\documentclass[12pt,preprint]{aastex}

\newcommand{\be}{\begin{equation}}
\newcommand{\ee}{\end{equation}}

\begin{document}
\title{Are Proxima and Alpha Centauri Gravitationally Bound?} 

\bigskip
\author{Jeremy G. Wertheimer} 
\bigskip
\author{Gregory Laughlin}

\bigskip
\bigskip
\bigskip
\bigskip
\affil{UCO/Lick Observatory, University of California,  Santa Cruz, CA 95064}

\begin{abstract} 
Using the most recent kinematic and radial velocity data in the literature, 
we calculate the binding energy of
Proxima Centauri relative to the center of mass of the $\alpha$ Centauri system.
When we adopt the centroids of the observed data, we 
find that the three stars constitute a bound system, albeit with a
semi-major axis that is on order the same size as $\alpha$ Centauri AB's Hill radius
in the galactic potential. We carry
out a Monte Carlo simulation under the assumption that the errors in
the observed quantities are uncorrelated. In this simulation, 44\% of the
trial systems are bound, and systems on the 1-3 $\sigma$ tail of the 
radial velocity distribution can have Proxima currently located near the apastron 
position of its orbit. Our analysis shows that a further, very significant
improvement in the characterization of the system can be gained
by obtaining a more accurate measurement of the radial
velocity of Proxima Centauri.
\end{abstract}

\keywords{stars: individual (Proxima Centauri, $\alpha$ Centauri)}

\section{Introduction} 
In 1839, the Scottish astronomer Thomas Henderson learned of
Bessel's successful parallax measurement of 
$0.314\arcsec \pm 0.020\arcsec$ for 61 Cygni \citep {Bessel1838}.
Bessel's result gave Henderson the confidence to report his own measurement
of a $1.16\arcsec \pm 0.11\arcsec$ parallax
for the $\alpha$ Centauri AB binary pair 
\citep{Henderson1839}\footnote{The Henderson (1839) reference is not 
returned by an
ADS reference search on Author=Henderson. However, a scanned copy of Henderson's
1839 MNRAS notice is available at:
http://articles.adsabs.harvard.edu//full/seri/MNRAS/0004//0000168.000.html}.
Henderson's distance determination of 0.86 pc for $\alpha$ Centauri
is reasonably close to the modern estimate of 1.35 pc. $\alpha$ 
Centauri remained the closest known stellar system until the
measurement of Proxima Centauri's parallax \citep{Voute17}.

Proxima and $\alpha$ Centauri exert a special fascination by virtue of
the fact that they are the Sun's closest neighbors. Their hold on the
imagination is further strengthened by the remarkable similarities of the 
masses and ages of $\alpha$ Cen A and B to the Sun.
Proxima lies a mere $15,000 \pm 700 \rm{AU}$ from the $\alpha$ Centauri 
binary pair,
and it has a small relative velocity, $\Delta V=0.53\pm0.14 \, {\rm km~s^{-1}}$
with respect to $\alpha$ Cen. The likelihood of such a stellar 
configuration occurring
purely by chance is less than $10^{-6}$ \citep{Matthews93}, and based
on this incredibly improbable arrangement, it has been suspected that
the stars constitute a bound triple system ever since Proxima's 
discovery \citep{Innes1915}.

The dynamics of Proxima and $\alpha$ Centauri have been most recently 
examined
by \citet*{Matthews93} and by \citet*{Anosova94}. Both
groups of authors showed that the (then-current) 
kinematic observations of the stars implied a positive total
energy for the three-star system. \citet*{Matthews93} sought to 
explain this somewhat disturbing result by hypothesizing that 
the measured masses of $\alpha$
Centauri AB were too low. They argued that a $3-\sigma$ increase in the 
dynamical mass of the system would result in a bound configuration.
\citet*{Anosova94} suggested that the three stars are members of a
``stellar moving group'', thus rendering Proxima's apparently hyperbolic 
trajectory less of an unusual occurrence.

In addition to the reasons presented by  \citet*{Matthews93} 
for the importance of understanding
the dynamical condition of the $\alpha$ Cen system, Proxima could also play a
role in volatile enrichment of any terrestrial planets that might be
orbiting either member of the central pair.
A study by \citet*{Wiegert97} showed that terrestrial planets are
dynamically stable when placed within 4 AU from either star of the $\alpha$
Centauri binary pair. Furthermore, accretion calculations by \citet{Lissauer04} 
suggest that terrestrial planet formation could have readily occurred
within the $\alpha$-Centauri system. One might therefore wonder about the 
habitability of putative terrestrial planets in the system. A possible 
concern with
respect to habitability arises because any planets orbiting the
$\alpha$-Centauri binary may be depleted in volatiles. If Proxima were bound
to the system during its formation stages, then it may have gravitationally
stirred the circumbinary planetesimal disk of the $\alpha$
Centauri system, thereby increasing the delivery of volatile-rich material
to the dry inner regions. 

\clearpage

\section{Data and Method} 
The Hipparcos satellite returned 
extremely accurate kinematic 
information on all three stars
of the $\alpha$ Cen system \citep{Hipp97}.  
We can combine these data with mass and 
radial velocity information to formulate a complete dynamical picture of 
the system (see Table~\ref{tbl-1}).

The kinematic data in Table~\ref{tbl-1} are consistent
with a broad spectrum of dynamical configurations for the three stars.
We wish to determine what fraction of these allowed configurations
correspond to bound systems. In order to do this, 
we perform an N=10000 Monte Carlo simulation by assuming that each observed 
parameter is independent and varies with a normal distribution implied by its
1-$\sigma$ error.  The only exception to 
the independence of the observations is that the parallaxes of $\alpha$ Cen A 
and B are set equal to each other within each realization, but 
still vary in a normal distribution from one realization to the next.  
We use observations of both $\alpha$ Cen A and B and the masses of 
each to calculate the properties of the center of mass of the A/B system.  
We calculate the gravitational binding energy $E_{tot}$ between the 
center of mass of the $\alpha$ Cen A/B binary and Proxima Centauri.  
We did this by converting to Cartesian coordinates and using cgs units 
with the simple formula
\begin{equation}
\label{eq:Etot}E_{tot}=T_{A/B}+T_{C}+U_{A/B/C}
\end{equation}
where $T_{A/B}$ and $T_{C}$ are the kinetic energies of the $\alpha$ 
Cen A/B center of mass and Proxima Centauri respectively and $U_{A/B/C}$ 
is the gravitational potential energy of the two point mass system.

\section{Results} 

\medskip We first calculated the properties of the $\alpha$ Cen A/B/C 
system using our method, but with the old observational data used by 
\citet{Anosova94}.  Following \citeauthor{Anosova94} we also restricted 
our errors 
in the N=1000 Monte Carlo simulation to within their 2-$\sigma$ values.
We find the A/B/C system is unbound with a probability 
of P=1.0 and that using the non-cgs (simulation) units from 
\citeauthor{Anosova94}, the energy of the system  was 5 $\pm$ 5 which 
was consistent with their results of P=1.0 and energy=6 $\pm$ 6.  
Using more accurate data from Hipparcos \citep{Hipp97}, \citet{Pourbaix02}, 
and  Queloz, D. (2004), not limiting the Monte Carlo errors to  
2-$\sigma$ and using cgs units in an N=10000 simulation, we find that 
the binding energy of the A/B/C system is now $5 * 10^{40} \pm 17* 10^{40}$ 
Joules with an unbound probability of P=0.55.  With modern data it is now
clear that almost half the realizations result in a bound triple system.
Using the centroid of each observation we can compute an estimate
of the observed orbital characteristics of Proxima Centauri with respect to
the center of mass of the A/B binary (see Table~2).

\begin{figure}
\plotone{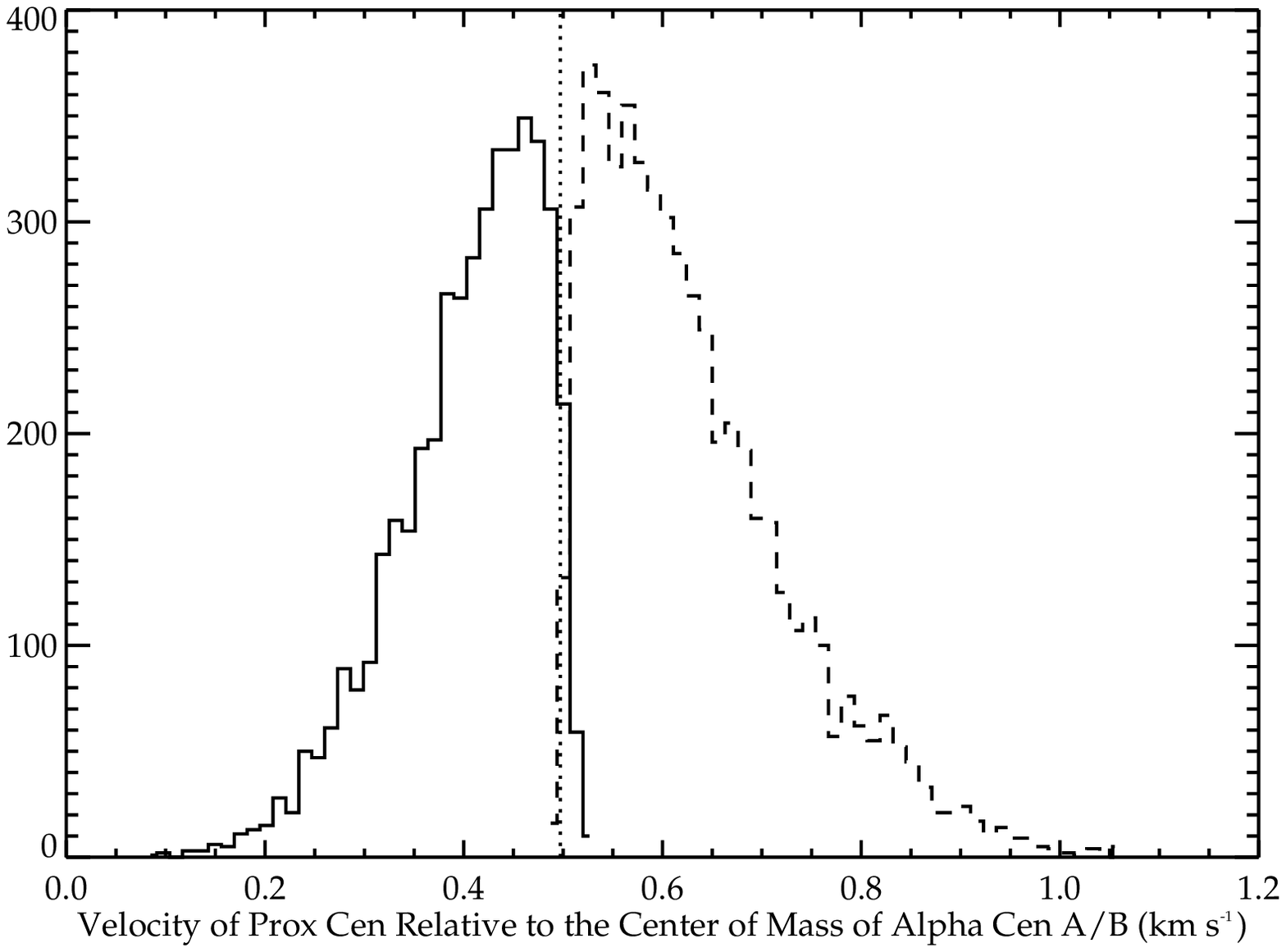}
\caption{\label{hist}Histogram of the number of bound and unbound 
orbits of Proxima Centauri around $\alpha$ Cen A/B in the Monte Carlo 
simulation.   The solid histogram represents the bound orbits while the 
dashed histogram represents the unbound orbits.  The dotted vertical 
line represents the centroid relative velocity, which yields a barely
bound orbit.}
\end{figure}

\begin{figure}
\plotone{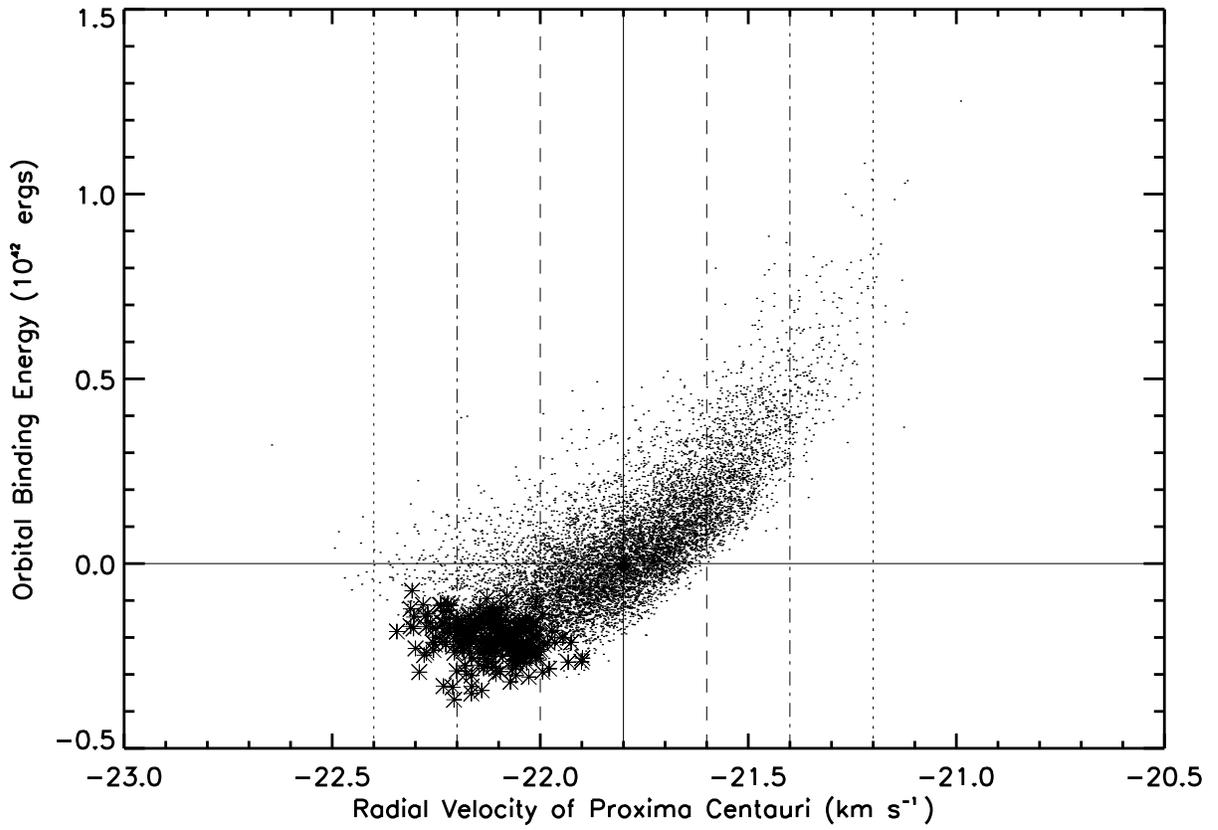}
\caption{\label{vr}Distribution of total system energy versus radial 
velocity of Proxima Centauri in each Monte Carlo realization.  The 
asterisks are orbits where Proxima Centauri is near apastron.  The 
diamond at the center is the centroid value from the observations.  
The vertical dashed lines represent the 1, 2, and 3-sigma deviations of 
radial velocity.}
\end{figure}

\medskip  For the case of a highly eccentric Keplerian orbit, Proxima 
Centauri would spend most of its time near apastron.  The current distance 
of Proxima Centauri from $\alpha$ Cen A/B is $15000\pm700$ AU, thus 
the estimated  semi-major axis of 272212 AU is almost certainly far too large.  
The velocity of Proxima Centauri relative to $\alpha$ Cen A/B in the 
Monte Carlo simulation is $0.53\pm0.14$ km~s$^{-1}$.   Systems with a 
lower relative velocity generally result in a bound orbit while higher 
velocities tend to result in a hyperbolic orbit (see Figure~\ref{hist}).  
According to observational data, Proxima Centauri is right on the cusp 
of being unbound, but we expect to find Proxima Centauri near apastron, 
which requires a slightly smaller relative velocity.  Only P=0.025 of 
the runs had a low enough relative velocity to situate Proxima Centauri 
fairly near apastron.  In Figure~\ref{vr}, we can see that the radial 
velocity of Proxima Centauri largely determines the energy of the resulting 
system and that a change of only 1 to 3-$\sigma$ in that radial velocity 
can result in a bound system currently near apastron.

\section{Conclusion}

\begin{figure}
\plotone{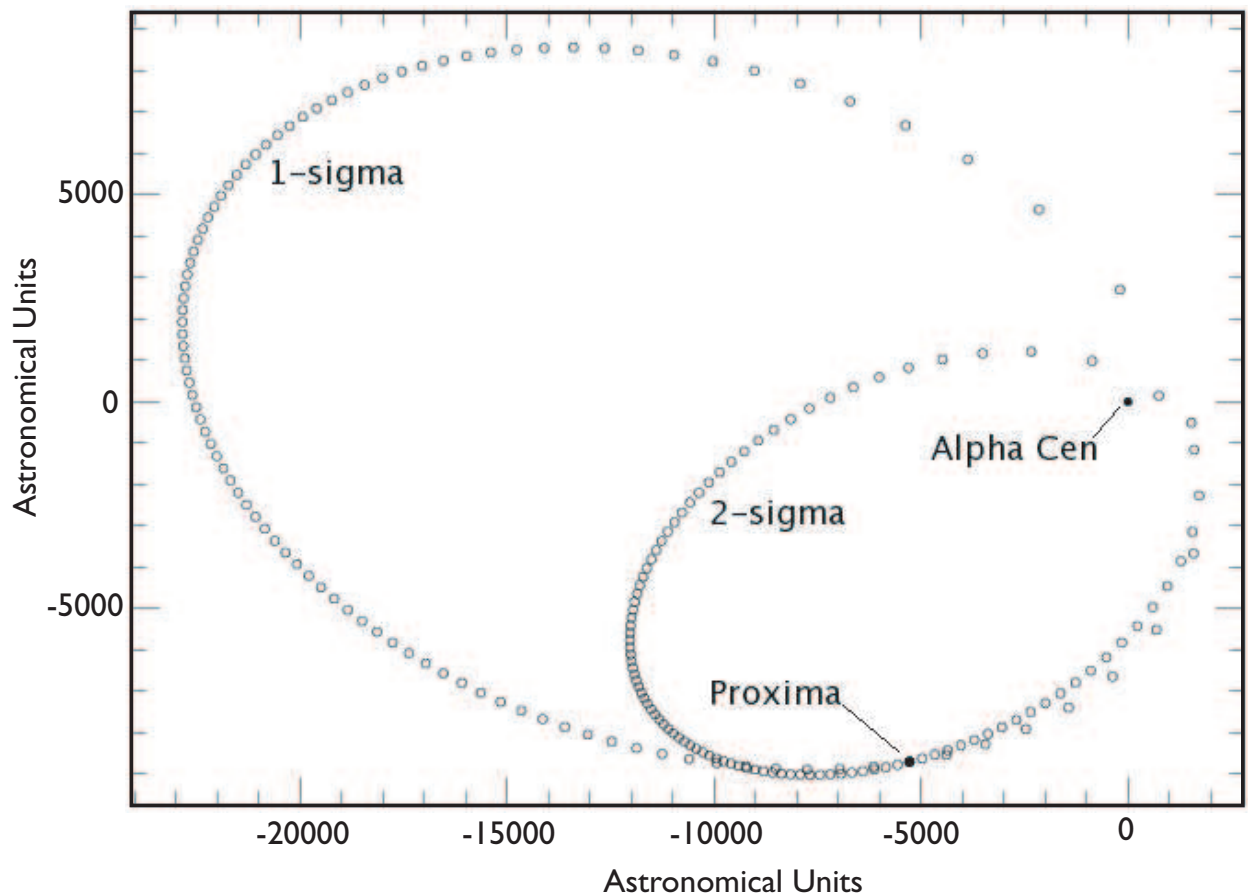}
\caption{\label{orbit}Two simulated orbits of Proxima Centauri around 
$\alpha$ Centauri in the plane of the sky as seen from Earth.  When 
Proxima Centauri has a radial velocity 1-$\sigma$ below the measured 
value, it results in a larger orbit while a 2-$\sigma$ radial velocity 
results in a smaller orbit.  Notice that in the 2-$\sigma$ orbit, Proxima 
is closer to apastron which we would expect.  The 0 and 3-$\sigma$ orbits 
are too large to be displayed on this scale.}
\end{figure}

The availability of Hipparcos data has provided us with the ability to
implement a significant improvement 
over previous studies of the Alpha Centauri system. 
Our results indicate that it is quite likely that Proxima Centauri is
gravitationally bound to the Alpha Centauri AB pair, thus suggesting that
they formed together within the same birth aggregate, and that they three
stars have the same ages and metallicities.
As future observations bring increased accuracy to the kinematic
measurements, it will likely become more obvious that
Proxima Centauri is bound to the $\alpha$ Cen A/B binary and that Proxima 
Centauri is currently near the apastron of an eccentric orbit.
(see Figure~\ref{orbit}).  Based on the expectation that the actual system is 
both bound and near apastron, we predict that improved measurements of
Proxima Centauri's absolute radial velocity will yield a value of 
-22.3 km s$^{-1} <v_{r}<-22.0$ km~s$^{-1}$.

{\bf Acknowledgment:} This work was supported by the 
US National Science Foundation CAREER 
Program under grant No. 0449986 to Greg Laughlin.

\clearpage

\clearpage

\clearpage

\clearpage

	\clearpage
	\newpage
	\bigskip

\begin{deluxetable}{lrrc}%left right right center aligned
\tablewidth{0pt}
\tablecaption{Observed Parameters and Masses for Proxima and $\alpha$ Cen A/B.\label{tbl-1}}
\tablehead{
\colhead{Parameter}		& \colhead{Proxima} &
\colhead{$\alpha$ Cen A/B\tablenotemark{a}}          &\colhead{ref}}
\startdata

$\alpha$ (RA) 	&$217^\circ.44894751\pm1.31$ mas 	&$219^\circ.91753275\pm51$ mas	&1\\
  	$\delta$   (Dec)                 		&  $-62^\circ.68135207\pm1.51$ mas       &  $-60^\circ.83712790\pm35$ mas          &1\\
	parallax (mas)                    		&  $ 772.33 \pm 2.42 $                       		&  $742.12 \pm1.4$			 		&1\\
	$\mu_{\alpha}$(mas yr$^{-1}$) &  $-3775.64\pm1.52$				& $-3642.53\pm12$                            		&1\\
	$\mu_{\delta}$(mas yr$^{-1}$)   &  $768.16\pm1.82$				&  $697.25\pm9$					&1\\
 	$V_R$ (km s$^{-1}$)                    &  $-21.8\pm0.2$    	&  $-22.445\pm0.0024$ 	&2,~3\\
 mass (M$_{\odot}$)          		&  $0.107\pm0.0214$               	&  $2.039\pm0.009$ 	&4,~3	\\ 

\enddata
\tablenotetext{a}{Errors in the $\alpha$ Cen A/B column are from the simulation (see text).}

\tablerefs{
(1) \citealt{Hipp97};
(2) Queloz, D., private communication 2004;
(3) \citealt{Pourbaix02};
(4) \citealt{Henry99}.
}
\end{deluxetable}
\clearpage

\newpage

\begin{table}
\begin{center}
\caption{Calculated Orbital Elements of Proxima Centauri around $\alpha$ Cen A/B.\label{tbl-2}}
\begin{tabular}{rrrrrr}
\tableline\tableline
  a (AU)         	& Ellipticity		& Inclination ($^\circ$)	&$\omega$ ($^\circ$) \tablenotemark{a}	&$\Omega$ ($^\circ$) \tablenotemark{b}	&Mean Anomaly ($^\circ$)\\
\tableline
 272212.148 	& 0.985	&150.9		&86.87	&197.81	&359.53\\

\tableline
\tablenotetext{a}{~argument of periastron}
\tablenotetext{b}{~position angle of the ascending node}
\end{tabular}
\tablecomments{These values are from the centroids of the observations and not our best guess from the Monte Carlo simulation.}

\end{center}
\end{table}


\begin{thebibliography}{}
\bibitem[Anosova, Orlov, \& Pavlova(1994)]{Anosova94} Anosova, J., Orlov, V.~V., \& Pavlova, N.~A. 1994, \aap, 292, 115

\bibitem[Bessel(1838)]{Bessel1838} Bessel, F. W. 1838, \mnras, 4,~152

\bibitem[ESA(1997)]{Hipp97} ESA. 1997, The Hipparcos and Tycho Catalogues, ESA SP-1200

\bibitem[Henderson(1839)]{Henderson1839} Henderson, T. 1839, \mnras, 4,~168

\bibitem[Henry(1999)]{Henry99} Henry, T. J., Franz, O. G., Wasserman, L. H., Benedict, G. F., Shelus, P. J., Ianna, P. A., Kirkpatrick, J. D., \& McCarthy, Jr., D. W. 1999, \apj, 512, 864

\bibitem[Innes(1915)]{Innes1915} Innes, R. T. A. 1915, Union Observatory Circular, 30 

\bibitem[Lissauer et al.(2004)]{Lissauer04} Lissauer, J.~J., Quintana, E.~V., Chambers, J.~E., 
Duncan, M.~J., \& Adams, F.~C. 2004, in Revista Mexicana de Astronomia y Astrofisica Conference Series, ed. G.~Garcia-Segura, G. Tenorio-Tagle, J. Franco, \& H.~W.~Yorke, 99-103

\bibitem[Matthews \& Gilmore(1993)]{Matthews93} Matthews, R. \& Gilmore, G. 1993, \mnras, 261, L5

\bibitem[Pourbaix et al.(2002)]{Pourbaix02} Pourbaix, D., Nidever, D., McCarthy, C., Butler, R.~P., 
	Tinney, C.~G., Marcy, G.~W., Jones, H.~R.~A., 
	Penny, A.~J., Carter, B.~D., Bouchy, F., Pepe, F.,  
	Hearnshaw, J.~B., Skuljan, J., Ramm, D., \& Kent, D. 2002, \aap, 386, 280

\bibitem[Vo{\^u}te(1917)]{Voute17} Vo{\^u}te, J. 1917, \mnras, 77, 650

\bibitem[Wiegert \& Holman(1997)]{Wiegert97} Wiegert, P.~A. \& Holman, M.~J. 1997, \aj, 113,1445
\end{thebibliography}
\end{document}